\documentclass[final,5p,times,twocolumn]{elsarticle}
\usepackage{graphicx}
\usepackage{amssymb}
\usepackage{lineno}
\usepackage{amssymb}
\usepackage[colorlinks,linkcolor=SB,citecolor=red]{hyperref}
\usepackage{geometry}
\usepackage{graphicx}
\usepackage{natbib}
\usepackage{etoolbox}
\usepackage{float}

\usepackage{hyperref}
\usepackage{float}
\usepackage[caption = false]{subfig}
\usepackage{amsmath,amssymb,amsfonts}

\apptocmd{\sloppy}{\hbadness 10000\relax}{}{}

\biboptions{comma,sort&compress}
\journal{Physica A}

\begin{document}

\begin{frontmatter}


\title{Scaling Features of Price-Volume Cross Correlation}


\author[ardalankiaaddress1,ccnsdaddress]{Jamshid Ardalankia}
\author[ardalankiaaddress1]{Mohammad Osoolian}
\ead{mohammadosoolian@gmail.com}
\author[emmanueladdress1]{Emmanuel Haven}
\ead{ehaven@mun.ca}
\author[ccnsdaddress,grjafariaddress1,grjafariaddress2]{G.Reza Jafari}
\ead{gjafari@gmail.com}

\address[ardalankiaaddress1]{Department of Financial Management, Shahid Beheshti University, G.C., Evin, Tehran, 19839, Iran}
\address[ccnsdaddress]{Center for Complex Networks and Social Datascience, Department of Physics, Shahid Beheshti University, G.C., Evin, Tehran, 19839, Iran}
\address[emmanueladdress1]{Faculty of Business Administration, Memorial University, St. John's, Canada and IQSCS, UK}
\address[grjafariaddress1]{Department of Physics, Shahid Beheshti University, G.C., Evin, Tehran, 19839, Iran}
\address[grjafariaddress2]{Department of Network and Data Science, Central European University, 1051 Budapest, Hungary}

\begin{abstract}
Price without transaction makes no sense. 
Trading volume authenticates its corresponding price, so there exists mutual information and correlation between price and trading volume. We are curious about fractal features of this correlation and need to know how structures in different scales translate information. 
To explore the influence of investment size (trading volume), price-wise (gain/loss), and time-scale effects, 
we analyzed the price and trading volume and their coupling by applying the MF-DXA method. 
Our results imply that price, trading volume and price-volume coupling exhibit a power law and are also multifractal.
Meanwhile, considering developed markets, the price-volume couplings are significantly negatively correlated.
However, in emerging markets, price has less of a contribution in price-volume coupling. 
In emerging markets in comparison with the developed markets, 
trading volume and price are more independent.

\end{abstract}


\end{frontmatter}



\section{Introduction}

The beginning of a comprehensive analysis of markets requires studying the markets' features 
with the consideration 
of some sort of scale-wise and magnitude-wise phenomena. 
This implies that the researcher needs to consider scaling behaviors of financial times series during the analysis. This  
was firstly done by Lux and Ausloos~\cite{Lux2002}.  Also, some scholars investigated the multifractal 
features~\cite{Peng1994,Peng1995,Shadkhoo2009,Hedayatifar2011,Ossadnik1994,
	Kantelhardt2002,Kavasseri2005}
of stock markets~\cite{Lux2002,Jiang2007,Podobnik2008,Podobnik2009,Lin2014,Sornette2018}, foreign exchange markets~\cite{Hedayatifar2011,Caraiani2015}, commodity markets~\cite{Zhuang2014}, cryptocurrency markets~\cite{Drod2018}, macro-economics time series~\cite{Safdari2016}, and financial multifractal network analyses~\cite{Torre2017}. Regarding to other scientific fields, some scholars have carried out fractal analysis on some phenomena in meteorology~\cite{Ausloos2004,ivanova2001multifractality}
	, text structure analysis~\cite{Ausloos2012}, hydrogeology~\cite{Ausloos2017}, and astronomy~\cite{Movahed2006}.

Among different variables in a market, the phenomenon which leads to price
discovery in an auction, is translation of trading volume. In this regard,
some scholars~\cite%
{Ausloos2002,Podobnik2009,Nasiri2018,Guo2012,Chen2001,Chuang2009,Campbell1993,Ahmad2016,Osborne1959,Saatcioglu1998,Wang1994}
investigated on the existence of information translation among price and
trading volume. Nevertheless, high trading volume may help one to
distinguish how much tendency there exists to a certain price. Hence, if a
certain price corresponds to a relatively high trading volume, it can be an
indication of high price reliability~\cite{Nasiri2018}, at that
time, in the market. It may lead to small spreads and also small liquidation
volatilities~\cite{Huang2018} within that time-scale. To clarify the
importance of investigating price-volume information together, Ausloos and
Ivanova~\cite{Ausloos2002} combined classical technical analysis with
thermodynamic proxies. Considering that from a physics point of view, it is
doubtful to investigate price movements without price-volume dependencies,
they obtained some fruitful measures which can model market behavior in
terms of trends and amplitude of volatilities by price and trading volume
time series. Since trading volume is a touchstone of how traders work in the
market and translate information via their transactions, this phenomenon can
shape movements in a market. 

To clarify the multifractal features of price-volume coupling, Podobnik 
\textit{et al.}~\cite{Podobnik2009} obtained a power-law in price-volume
cross correlations. They investigated logarithmic changes of price and its
corresponding volume and showed that just the cross correlation of absolute
values of volatilities are statistically significant. It is noteworthy to
state that in the developed markets, the entanglement of price return
fluctuations and trading volume~\cite{Nasiri2018}, leads to a
situation whereby an increase in volume shall --to some extent as a turning
point-- increase price fluctuations. But then, a further increase in trading
volume leads to lowering the price fluctuation. Meanwhile, Guo~\textit{et al.%
}~\cite{Guo2012} investigated the price-volume cross-correlation of
commodities in futures markets and compared their multifractal behaviors. %

In reference to the theme of this current paper, Podobnik~\textit{et al.}~%
\cite{Podobnik2009}; Nasiri~\textit{et al.}~\cite{Nasiri2018} and Guo 
\textit{et al.}~\cite{Guo2012}, found that price-volume cross volatilities,
may have scaling behavior at the levels of time and magnitude-scales. Since
it makes no sense to study price volatilities without considering the
reliability which a certain price owes to trading volume, we investigate the
scaling behavior of the cross correlation of price-volume volatilities. 
Different scaling behavior toward events of different magnitudes, causes
multifractal behavior~\cite{Hedayatifar2011}.

In order to reveal scaling behavior, one needs to differentiate large-scale
and small-scale patterns in a stock market as complex system~\cite%
{Ossadnik1994,Peng1995,Buldyrev1995,Barnes1966}. Since large-scale patterns
(main trends) are somehow evident they do not present much latent
information~\cite{Hedayatifar2011}. Hence, initially it is necessary to
apply ``detrended fluctuation analysis'' (DFA) for non-stationary signals.
See Peng \textit{et al.}~\cite{Peng1994} and also~\cite%
{Ossadnik1994,TAQQU1995,Kantelhardt2001,Hu2001,Chen2002}.

Because of psychological biases and irrational decisions of investors~\cite%
{Kwapie2012,Chen2016,Pece2015,Ozturk2017,Hong2016,Chan2015}; out-of-market
effects~\cite{Kwapie2012,Dash2016}; uncertainty about the transparency of
fundamental analysis~\cite{Chen2016,Abarbanell1997}; the coexistence of
collective effects and noise~\cite{Kwapie2012,Jamali2015}; and the lagging
diffusion of internal and external information between several dynamics, a
situation occurs where some assumptions of the EMH (Efficient Market
Hypothesis) become doubtful. In spite of the complexity aspects in financial
markets such as self-organizing (in order to increase adaptability) and
scaling patterns of nonlinear dynamics, we observe power-law (scale
invariance) behavior in some scales~\cite{Kwapie2012,Tahmasebi2015}.
Industrial, economic and political cycles which are dynamic and ever
changing, may cause a variability of statistical properties of time series
in different scales. This is what we call `non-stationarity'. These
behaviors cause multifractal correlations between time series. Systems which
contain persistent nonlinear interactions as inputs and outputs of the
constituents, may cause some simultaneous and some lagging effects to
emerge. In the case of persistent information, these effects cause
long-range auto-correlation (for one constituent) and long-range
cross-correlation (between several constituents)~\cite%
{Buldyrev1995,Kantelhardt2001,Podobnik2011,Zhu2018}.

Since real-world financial time-series may be
non-stationary --possess some trends-- and may be limited in length, we need
a method which considers limited-length effects and non-stationarity effects~%
\cite{Movahed2006}. The mentioned methodology has been applied successfully
in finance~\cite%
{Lux2002,Podobnik2009,Carbone2004,Mantegna2000,Liu1999,Vandewalle1999,Yin2013,Zunino2008,Safdari2016,Zhu2018}
and has been developed to Fourier-DFA and MF-DFA~\cite{Kantelhardt2002}.

Besides these references, Podobnik and Stanley~\cite%
{Podobnik2008} introduced detrended cross correlation analysis (DXA) for
studying two power-law non-stationary time series. Then, Zhou~\cite{Zhou2008}
developed the DXA method to multifractal DXA (MF-DXA) to investigate
multifractal behaviors of two power-law non-stationary time series.
Moreover, to investigate coupling behavior among more than two
non-stationary power-law time series, some researchers introduced
Coupling-DXA method~\cite{Hedayatifar2011}. They believed that studying more
than two signals in complex systems leads researchers to better
understanding the FOREX market structure, and they showed that several (more
than two) constituents have a scaling coupling behavior. In the case of
combining the MF-DXA within magnitude-wise scales, several researchers~\cite%
{Shadkhoo2009,Cottet2004,Podobnik2009,Campillo2003,Hajian2010} have done
some efforts on detrended covariance.

Also, some scholars specifically cast light onto the detrending methods in
financial markets. As a heuristic method, Caraiani and Haven~\cite%
{Caraiani2015} applied EMD (Empirical Mode Decomposition) method on the
detrending process of MF-DFA. They investigated multifractality in the
currencies and non-linearity of the market. 
In this paper, we apply multifractal detrended cross correlation analysis (MF-DXA) 
to investigate the scaling behavior of volatilities of price-volume coupling in markets such as the 
DJIA, S\&P500, TOPIX, TSE and SSEC.  
It is shown that, the price and trading volume and their cross correlations, contain multifractal behaviors. 
Moreover, the correlation coefficients of their volatilities decrease with an increase of the time-scale. 
In section~\ref{Methodology}, a review on methodology is presented. Then in section 3, we analyze and describe the input data 
used in the methodology. 
In section 4, the empirical results are discussed and then in section 5, we provide for the conclusion.%
\section{Methodology}
\label{Methodology}
\paragraph{MF-DXA}
The general steps of MF-DXA methodology~\cite%
{Lux2002,Ossadnik1994,Peng1994,TAQQU1995,Kantelhardt2001,Hu2001,Chen2002}
are as follows. Initially we convert the time series to standardized
logarithmic changes. After demeaning each data-point, we execute a
cumulative summation which is called profile series (Eq.~\ref{eq1}). 
\begin{eqnarray}
X_{(i)} &=&\sum_{k=1}^{i}[x_{k}-<x>]  \nonumber \\
Y_{(i)} &=&\sum_{k=1}^{i}[y_{k}-<y>]  \nonumber \\
i &=&1,2,....,N.  \label{eq1}
\end{eqnarray}
Where $<...>$ is the average value of the time series; $N$ is the total
length; and $X_{(i)}$ and $Y_{(i)}$ are profile series. To investigate
time-scale effects on series, we consider several time-scale windows with
the length of $s$ which are iterated on the time series. Variable $N_{s}$ is
obtained by $int(\frac{N}{4})$ which provides the maximum number of segments
(with length of $s$) in each time series. Since the length $N$ may not be an
integer multiple of scale length $s$, we repeat the same iteration from
ending data point to starting one. It means that each non-overlapping scale
window, is allocated to sequential time locations without ignoring any data
points (Eq.~\ref{eq3}). Consequently, the total number of time-wise scales
would be $2N_{s}$.

In order to convert profile series to non-stationary type series, we should
eliminate the main trends (large-scale behaviors) from the time series.
Among lots of detrending methods (such as polynomial detrending, Fourier
detrending), it is better to start from the simplest one to avoid
over-fitting and also to avoid eliminating excessive amount of information.
So in this study, one-degree linear detrending is applied. As follows,
detrending covariance for each time-wise scale ($s$) and windows location ($%
\nu $) is obtained by subtracting the local trend of each window. After
detrending, local deviation in each non-overlapping window is obtained. 
\begin{equation}
N_{s}=int(\frac{N}{s})  \label{eq2}
\end{equation}

\begin{eqnarray}
F_{(s,\nu )}^{2} &=&\frac{1}{s}\sum_{j=1}^{s}|X_{(\nu -1)s+j}^{(j)}-\tilde{X}%
_{\nu }^{(j)}||Y_{(\nu -1)s+j}^{(j)}-\tilde{Y}_{\nu }^{(j)}|  \nonumber \\
\nu  &=&1,2,..,N_{s}  \nonumber \\
F_{(s,\nu )}^{2} &=&\frac{1}{s}\sum_{j=1}^{s}|X_{N-(\nu -N_{s})s+j}^{(j)}-%
\tilde{X}_{\nu }^{(j)}||Y_{N-(\nu -N_{s})s+j}^{(j)}-\tilde{Y}_{\nu }^{(j)}| 
\nonumber \\
\nu  &=&N_{s}+1,N_{s}+2,...,2N_{s};  \label{eq3}
\end{eqnarray}%
where $\tilde{X}_{\nu }$ and $\tilde{Y}_{\nu }$ are local trends which are
fitted polynomials in each window $\nu $ with the local length of $s$.%
\newline
For small (large) scale windows, the detrended covariance will be affected by small and rapid 
(large and slow) fluctuations. 
Because of the relatively short (long) scales, on average, 
there will be an increase (decrease) of local effects in the detrended covariance function. 
Accordingly, small-scale (large-scale) behaviors are observed significantly in small (large) windows. 
We can further explain this by saying that monofractals are normally distributed~\cite{Shayeganfar2009}
and the volatilities can be explained just by the second statistical moment, i.e. the `variance'. On the contrary, %
when it comes to multifractals, in large (small) scales, local variations
are excessively large (small). As a result, to consider different behaviors
of fluctuations, we magnify our concentration from small and frequent, to
large and rare fluctuations by weighting them. Hence, for considering
effects of events on magnitude scales, the parameter $q$ is applied. %
When a big $q$ is considered, 
a high weight is applied to the tails (rare and enormous events) of the logarithmic changes histogram.%
To be unbiased toward small or large variations, $q=0$ is applied. 
\[
F_{q}(s)=[\frac{1}{2N_{s}}\sum_{\nu =1}^{2N_{S}}{F^{2}}(s,\nu )^{q/2}]^{%
	\frac{1}{q}}\qquad q\neq 0.
\]%
%
\[
F_{q}(s)=exp{\{\frac{1}{4N_{s}}\sum_{\nu =1}^{2N_{s}}\ln [F^{2}(s,\nu )]\}}%
\qquad q=0.
\]%
%
In the case of the second moment, $F_{q}(s)$ leads to~$\sigma =\sqrt{\sigma_{1}\sigma _{2}}$~\cite{Hedayatifar2011,Podobnik2008}.\newline
If the time series are scale-invariant and have long-range correlation, the
MF-DXA approach would contain a scaling behavior as follows: 
\[
F_{q}(s)\sim s^{h_{xy}(q)},
\]%
%
where $F_{q}(s)$ is a fluctuation function in order of $q$ and scale length
of $s$. \newline
The slopes of $\log (F)-\log (s)$ are called the generalized Hurst exponent $%
h(q)$. For each moment $q$, there is an $F(s)$. Whilst within a range of
time-scales, the slope of $log(F)-log(s)$ for various moment $q$ is
constant, the time series is scale-invariant. On the contrary, if there is a
change in the value of slope $h(q)$ for a single $q$, it is called a
`cross-over' and the time series is now scale variant. For a range of $q$, a
spectrum of $h(q)$ is obtained. The degree of multifractality~\cite%
{Schumann2011} as a risk measure~\cite{Zhuang2014} is as follows: 
\[
\Delta h=h_{max}(q)-h_{min}(q).
\]%
%
If $\Delta h=0$, the system is monofractal so the time series does not have
segments with extreme small and extreme large fluctuations. Hence, its
detrended covariance within a same time-scale window length, when powered by 
$q$th-order in different windows location $\nu $, will yield no peak. On the
contrary, for $\Delta h\ne 0$ , the system is multifractal~%
\cite{Jafari2007}. In this case, the average value for residuals of local
trends for different $s$ and $\nu $, are not similar. So, large variations
dominate the results for a relatively large $q$, and small
variations dominate the results for a relatively small $q$%
. \newline
%
It is worthy to state, that if the Hurst
exponent $h^{Price}_{(q=2)}=0.5$, the time series is a random walk.
For a Hurst exponent $h^{Price}_{(q=2)}<0.5$ $(h^{Price}_{(q=2)}>0.5)$, the time series tends to be
anti-persistent (persistent) and is negatively-correlated
(positively-correlated). 
Also, the Hurst exponent values are as following: for the developed markets $0<H_{Price}<0.5$, and for the emerging markets $0.5<H_{Price}<1$%
. R\'{e}nyi's exponent (scaling exponent function) is as follows~\cite%
{Shadkhoo2009}: 
\[
\tau _{xy}(q)=qh_{xy}(q)-1.
\]%
%
If $\tau _{xy}(q)$ (which is a derivative of $h_{xy}(q)$) is a linear
function of $q$, the time series is monofractal. To examine the singularity
content of time series, the singularity spectrum is as follows: 
\begin{eqnarray}
\alpha  &=&h_{xy}(q)+qh_{xy}^{\prime }(q)  \nonumber \\
f_{xy}(\alpha ) &=&q(\alpha -h_{xy}(q))+1;  \label{eq9}
\end{eqnarray}%
where $\alpha $ is the singularity of a time series and $f_{xy}(\alpha )$ is
the multifractality spectrum. In addition, the multifractality strength is
observed by the singularity width $\Delta \alpha $, as below: 
\begin{equation}
\Delta \alpha =\alpha _{max}-\alpha _{min};  \label{eq10}
\end{equation}%
%
where $\alpha _{max}$ relates to $q_{min}$ and $\alpha _{min}$ relates to $%
q_{max}$ . If $\Delta \alpha =0$, the time series is monofractal and the
response of the cross correlation toward different events of $q$, in large
and small lengths of the time-scale is identical, and the multifractality
spectrum is just a point. \newline
Until this section, the fluctuation function $F(s)$ for each $q$, is
obtained for the coupling of time series.
\paragraph{MF-DFA}
When applying the detrended covariance function for just a single time
series, the detrended variance function is presented. The rest of the
methodology is similar to the MF-DXA. This process needs to be applied for
price and trading volume separately.
\paragraph{Correlation}
The absence of any correlation leads to $\rho _{DXA}=0$. Based on Podobnik
et al.~\cite{Podobnik2011} this claim is just valid for an unlimited-length
of time series. In the case of limited length, the situation of $\rho
_{DXA}\neq 0$, but no correlation is probable. If the investigated time
series have power law -as we will show in the rest of this research- with
the help of cross correlation statistic~\cite%
{Podobnik2011,Yin2013,Zhuang2014}, by the equation below we will show the
scales which the cross correlation coefficient is statistically significant: 
\begin{equation}
Q_{cc}(m)=N^{2}\sum_{i=1}^{m}\frac{X_{i}^{2}}{N-i},  \label{eq11}
\end{equation}%
%
where $X_{i}$ is cross correlation function as follows: 
\begin{equation}
X_{i}=\frac{\sum_{k=i+1}^{N}x_{k}y_{k-i}}{\sqrt{\sum_{k=1}^{N}x_{k}^{2}} 
	\sqrt{\sum_{k=1}^{N}y_{k}^{2}}};  \label{eq12}
\end{equation}%
%
where $x_{k}$ and $y_{k}$ are detrended logarithmic changes of time series. 
\newline
Since the cross correlation statistic $Q_{cc}(m)$ is somehow similar to the $%
\chi ^{2}(m)$ (Chi-squared) distribution, the critical value is measured by $%
\chi ^{2}(m)$. So, when $Q_{cc}(m)$ is less than the critical value (null
hypothesis) it means there is no significant cross correlation. On the
contrary, if $Q_{cc}(m)$ is more than the critical value (alternative
hypothesis), then there exists no reason to reject the existence of cross
correlation.\newline
In order to reveal cross-correlation scaling behavior, we calculate the
correlation by DXA fluctuation as follows: 
\begin{equation}
\rho _{DXA}=\frac{F_{DXA_{price-volume}}^{2}(s)}{%
	F_{DFA_{price}}(s)F_{DFA_{volume}}(s)};  \label{eq13}
\end{equation}%
%
where $\rho _{DXA}$ stands for the cross-correlation 
coefficient of price fluctuation function and the trading volume
fluctuation function and $s$ is the time-scale length. $%
F_{DFA}$ is the fluctuation function of each time series individually (price
or trading volume) and $F_{DXA}$ is the fluctuation function of
cross-correlation of price-volume.
\section{Empirical Data}
In this study, we gathered the price index and trading volume of DJIA,
S\&P500, TOPIX, TSE and SSEC markets during March $21^{st}$, 2013 until
March $20^{th}$ ,2018. This includes around 1300 trading days from which we
can investigate the multifractal behavior of time series volatilities and
their related cross correlations. 
\begin{center}
	\begin{figure}[ht]
		\centering
		\includegraphics[width=0.5\textwidth]
		{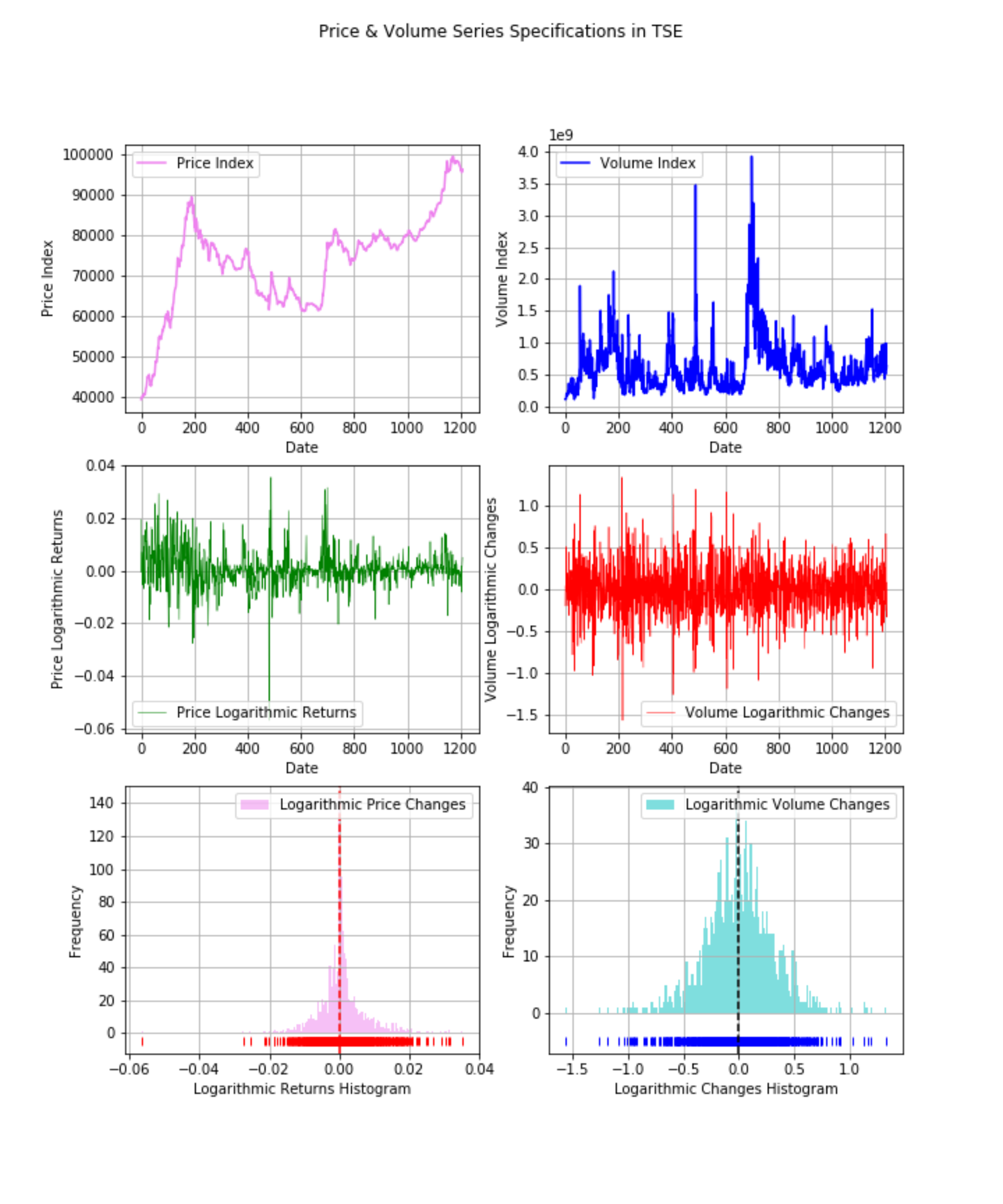}  \includegraphics[width=0.5%
		\textwidth] {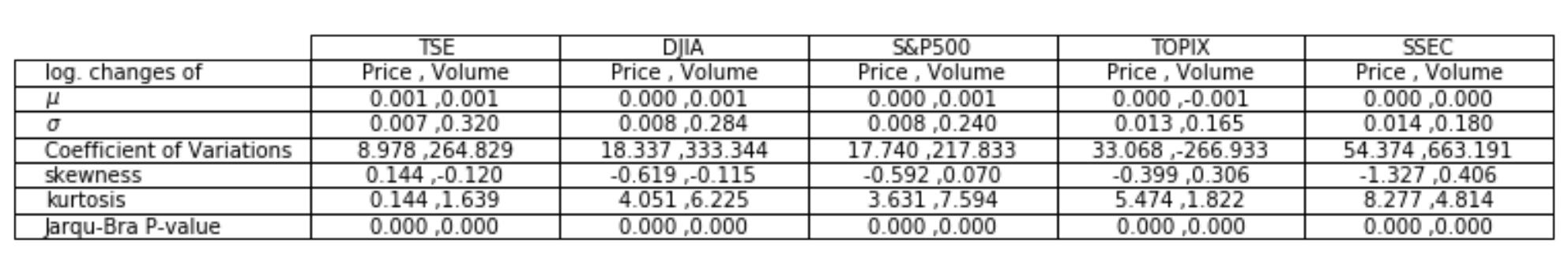}   
		\caption{Descriptive statistics of investigated markets are demonstrated. }
		\label{figSTAT}
	\end{figure}
\end{center}
In Fig.\ref{figSTAT} top panel, the descriptive statistics for one of the
markets is shown. Also, as illustrated in Fig.\ref{figSTAT} bottom panel,
none of the time series are normally distributed. Hence, they are not
monofractal~\cite{Shayeganfar2009}. Initially, after calculating the
logarithmic changes of time series, they were standardized. Then 
\begin{eqnarray}
r_{price} &=&\ln {P_{t}}-\ln {P_{t-1}}  \nonumber \\
r_{volume} &=&\ln {V_{t}}-\ln {V_{t-1}},  \label{eq14}
\end{eqnarray}%
%
where $t$ is daily data point, and also, $P$ and $V$ refer to price and
trading volume, respectively. The above mentioned results are used as inputs
to the methodology.
\section{Empirical Results}
\begin{figure}[t]
	\centering
	\includegraphics[width=0.5\textwidth]{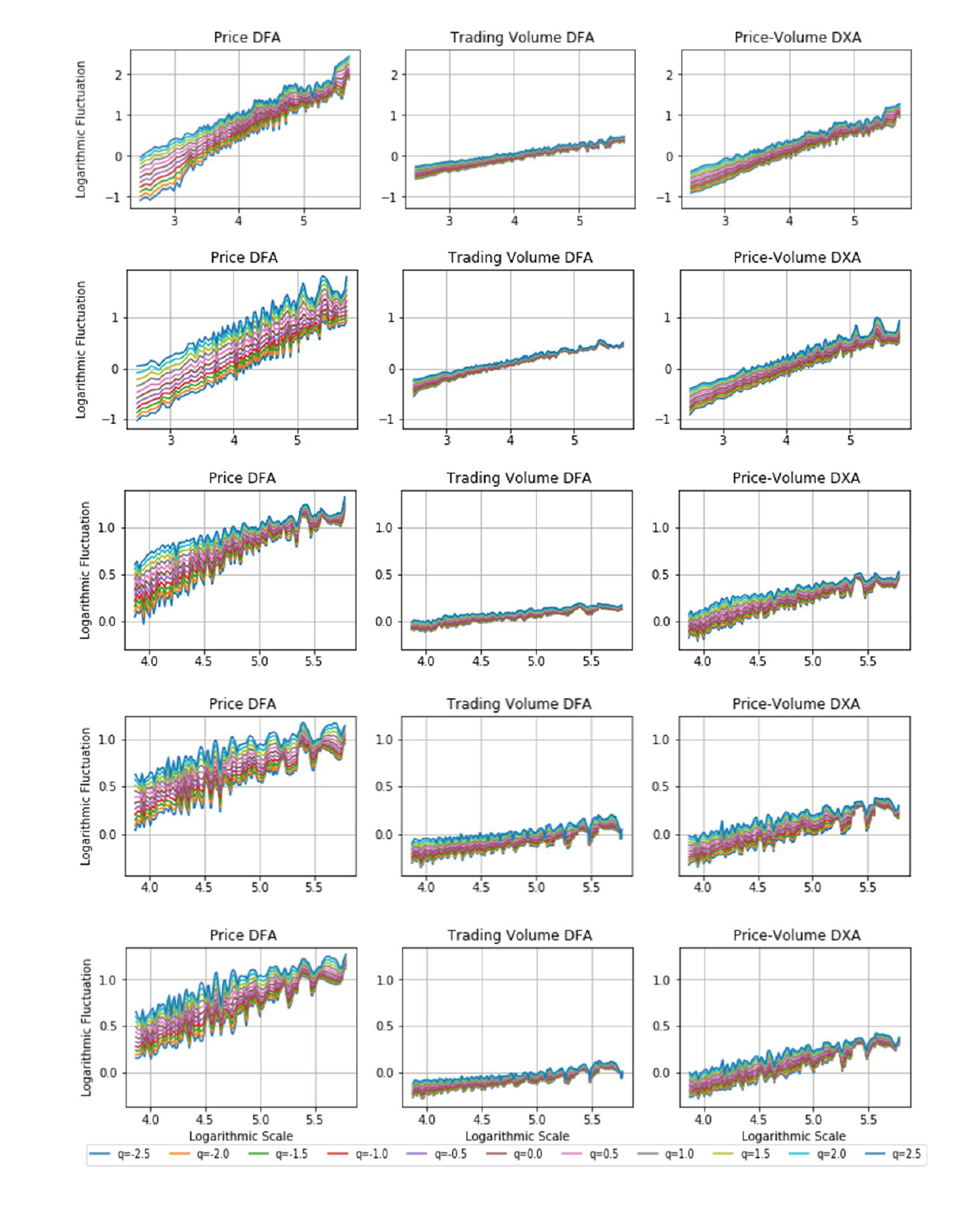}
	\caption{$log(F)-log(s)$ for price index (left) and trading volume (middle)
		and their cross correlation (right) are demonstrated. From top to bottom, $%
		log(F)-log(s)$ for TSE, SSEC, TOPIX, S\&P500, DJIA, are shown.}
	\label{logFlogS}
\end{figure}

\begin{figure}[ht]
	\centering
	\includegraphics[width=0.49\textwidth] {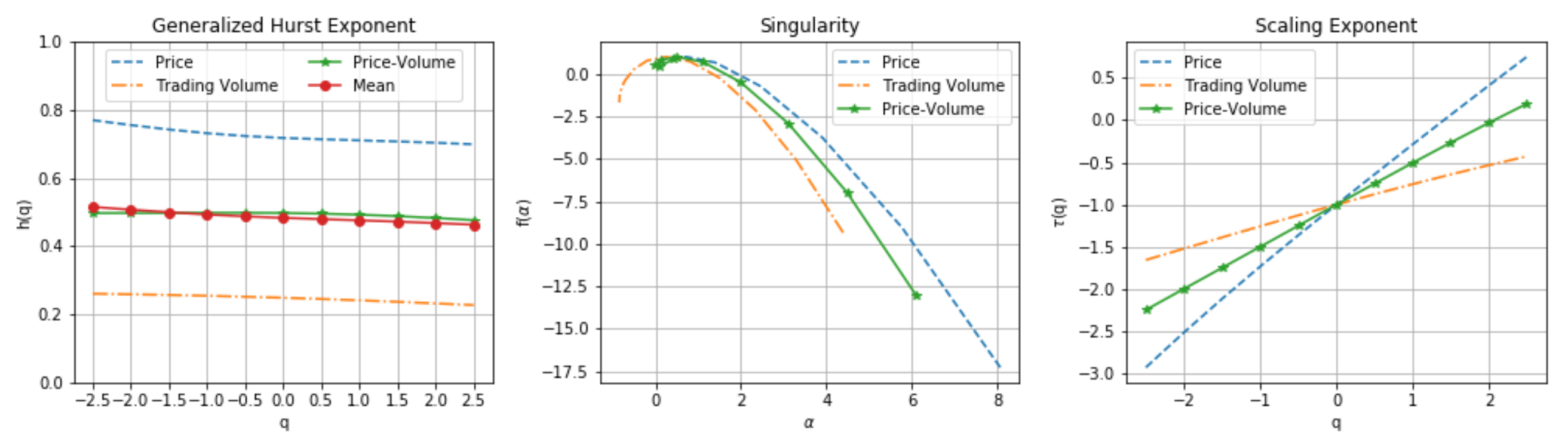}	
	\includegraphics[width=0.49\textwidth] {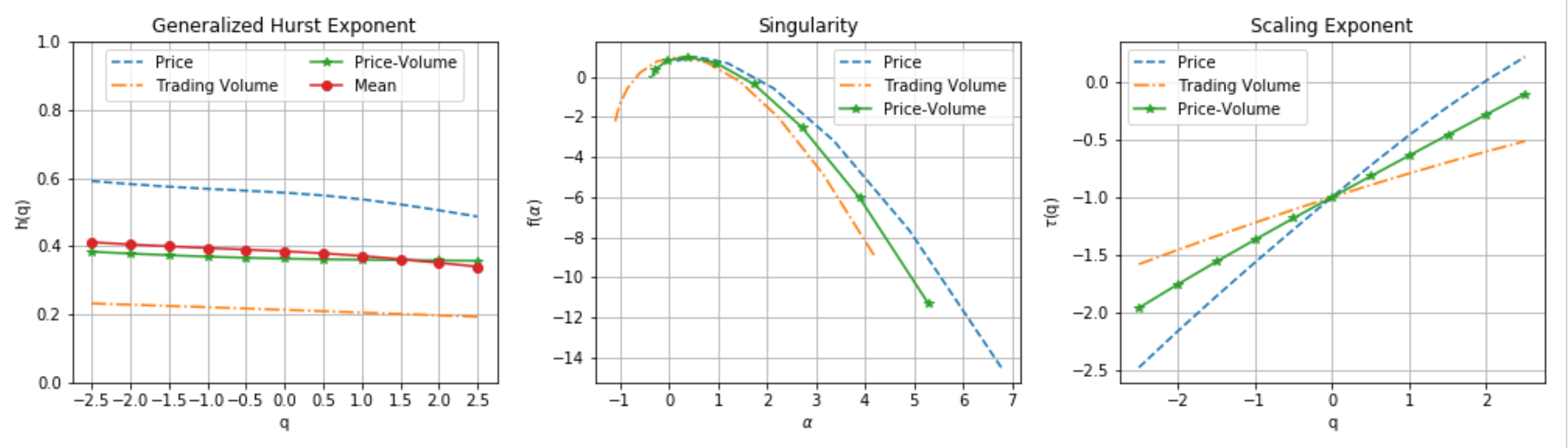}
	\includegraphics[width=0.49\textwidth] {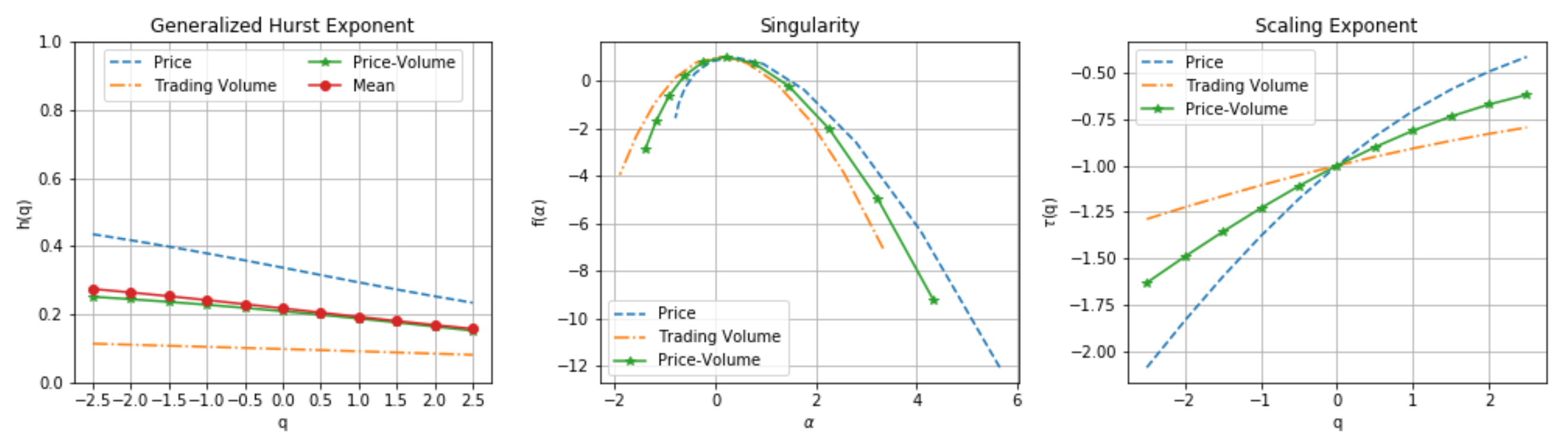}	
	\includegraphics[width=0.49\textwidth] {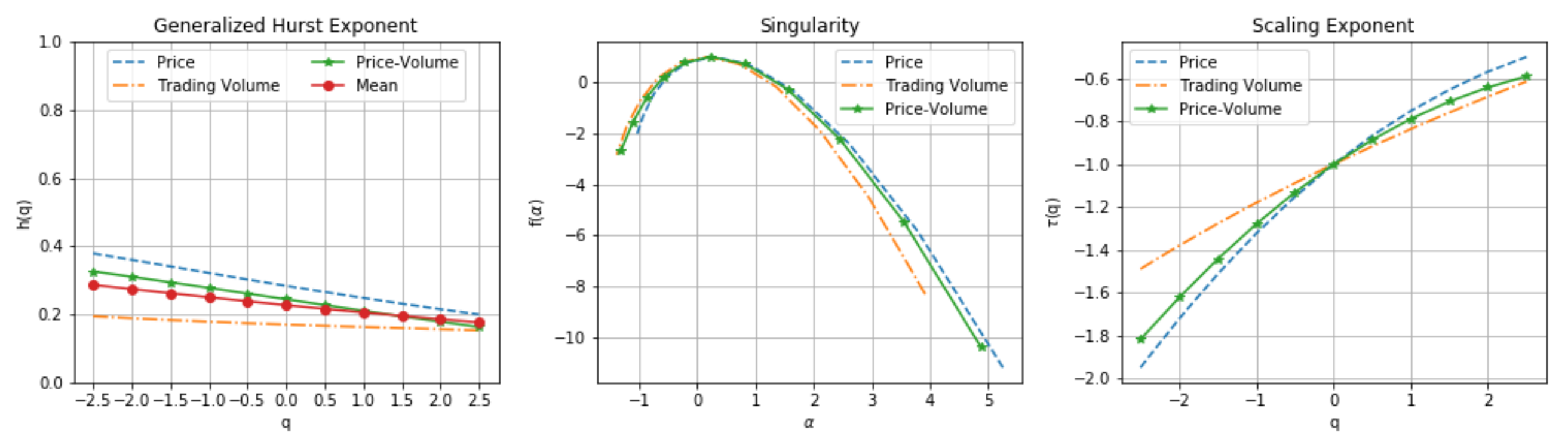}	
	\includegraphics[width=0.49\textwidth] {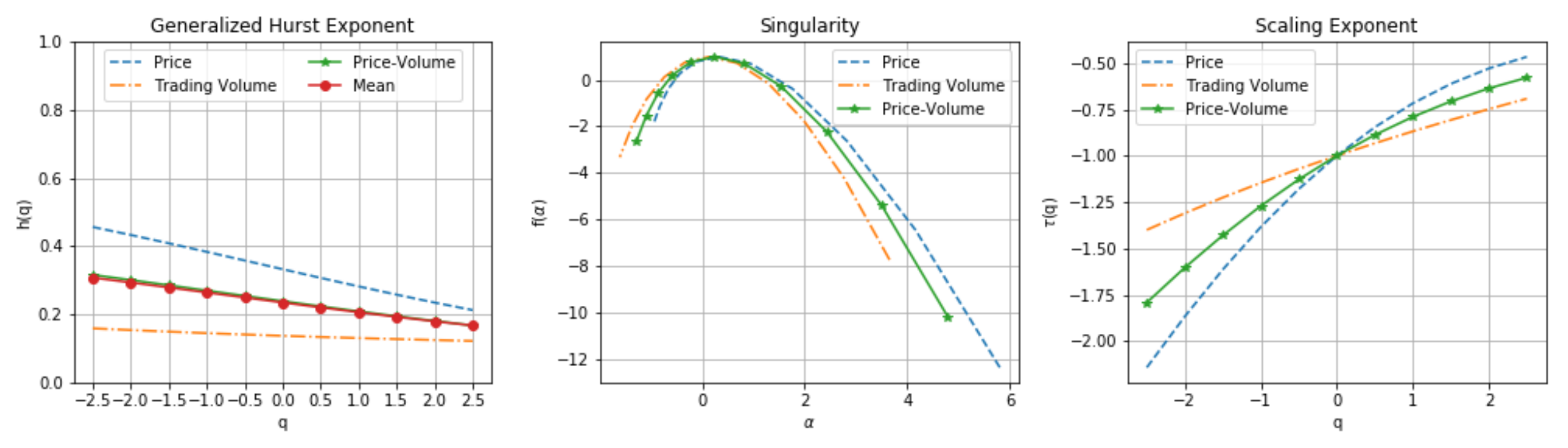}
	\caption{Multifractality features such as Hurst exponent spectrum for price index (left) and trading volume (middle) and their price-volume correlation (right) are demonstrated. From top to bottom, multifractality features of TSE, SSEC, TOPIX, S\&P500, DJIA, are shown.}
	\label{fig_MF}
\end{figure}

\begin{figure}[htbp]
	\centering
	\includegraphics[width=0.5\textwidth]
	{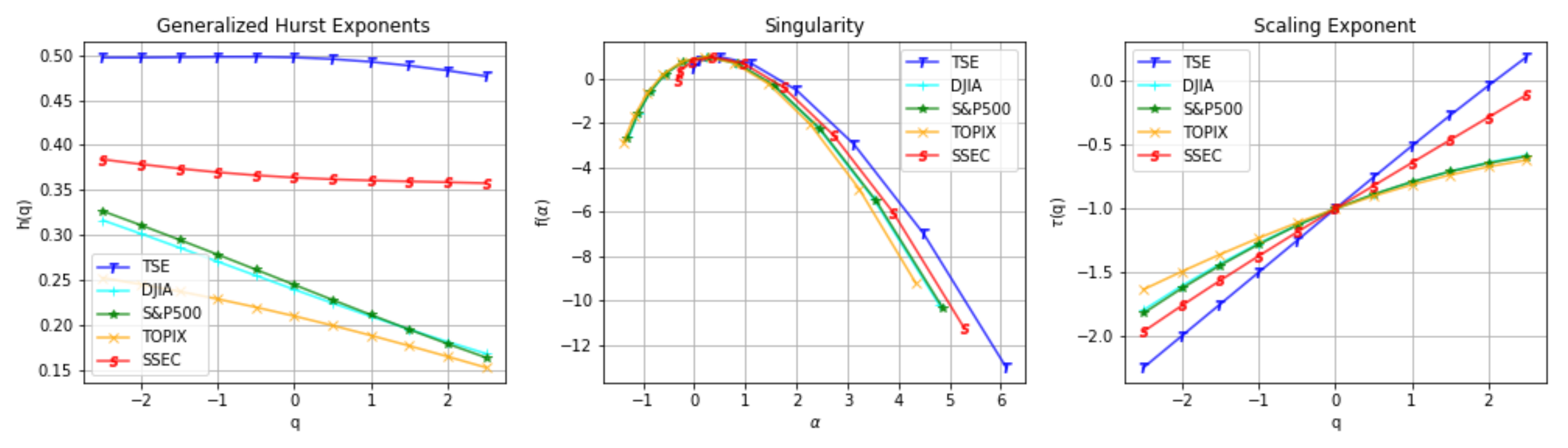}  \includegraphics[width=0.5%
	\textwidth] {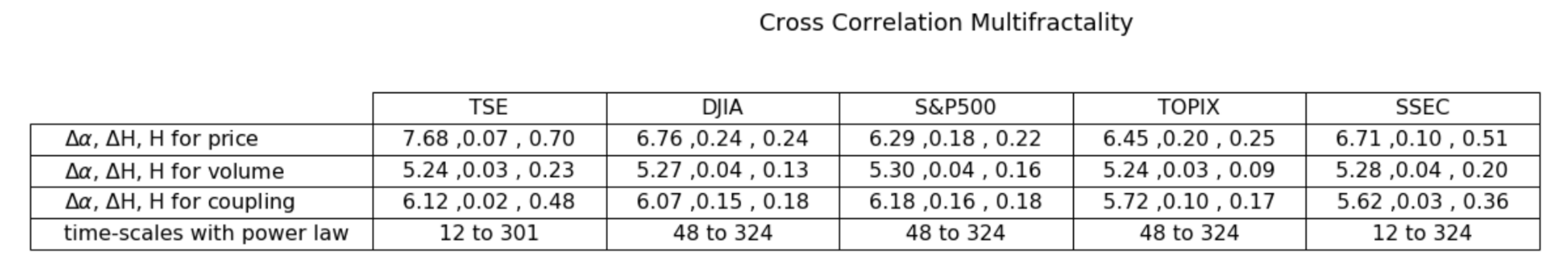}   
	\caption{\textit{Top:} Multifractality features of price-volume coupling:
		Hurst exponent spectrum (left), singularity spectrum (middle) and scaling
		exponent (right) are shown. \textit{Bottom:} Multifractality features of
		price time series, volume time series, and price-volume coupling are
		presented.}
	\label{fig_MFcross}
\end{figure}

\paragraph{MF-DXA}
After detrending in each non-overlapping segments, $\nu$, with length of $s$
for price, volume, and price-volume cross series, we shall in order to
estimate the power-law relation, compute $\log(F)-\log(s)$. This is
illustrated in Fig.~\ref{logFlogS} for the markets we investigated. In Fig.~%
\ref{fig_MF} and Fig.~\ref{fig_MFcross}, the multifractality features of
price-volume coupling for the investigated markets are presented. 

Altering a power law relation in different time-scales, means there exists scale 
variance and it is an indication different power-laws are needed to describe the corresponding time-series 
in different time-scales. On the other hand, a system is scale invariant
when it generates its structure in different intervals. To obtain valid
power-laws, proper scales are extracted which are shown in Fig.~\ref%
{fig_MFcross}. It is worthy to say that for window lengths near $N$, the
local trend of the segments become more similar to the whole time series %
rather than the too small segments, and in this process $%
F(s)$ becomes relatively independent of iterations on $\nu 
$.\newline
Another reason for this phenomenon is the effect size of variations. Large
segments (small segments) contain large-scale (small-scale) behaviors of the
time series. Hence, applying the $q$th-order effect on the fluctuation
function, leads to a divergence of $\log (F)-\log (s)$ for small scales.
Nevertheless, for $s\ll N$ (local deviation of detrended segments are small
on relatively small scales), the fitting polynomial %
--a linear local trend in this study-- is well fitted on
the segments and results in a magnification of the divergence of $F(s)$ for
different $q$ values.\newline
The generalized Hurst exponent presents multifractality of correlations.
Based on Podobnik and Stanley~\cite{Podobnik2008}, the Binomial measure from
the $p$-model for the cross-correlation exponent of the fractionally
auto-regressive integrated moving average (FARIMA) with identical stochastic
noises, is equal to their arithmetic mean at $q=2$. However, Zhou~\cite%
{Zhou2008} illustrated that the same relation for $q$ values other than 2,
is valid~\cite{Zhuang2014} which is presented in red color
in the left panels in \ref{fig_MF}. 

\begin{figure}[tp]
	\includegraphics[width=0.5\textwidth] {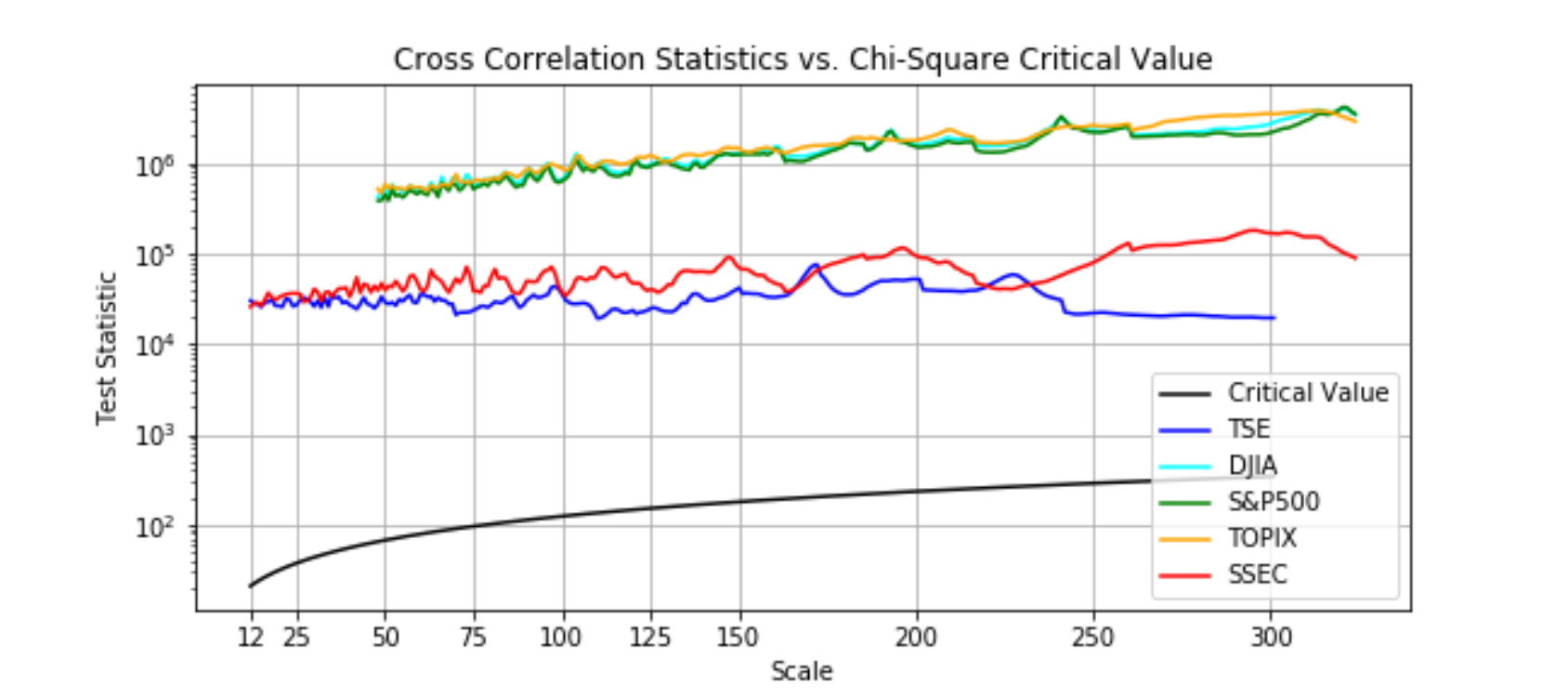}  
	\includegraphics[width=0.5\textwidth]
	{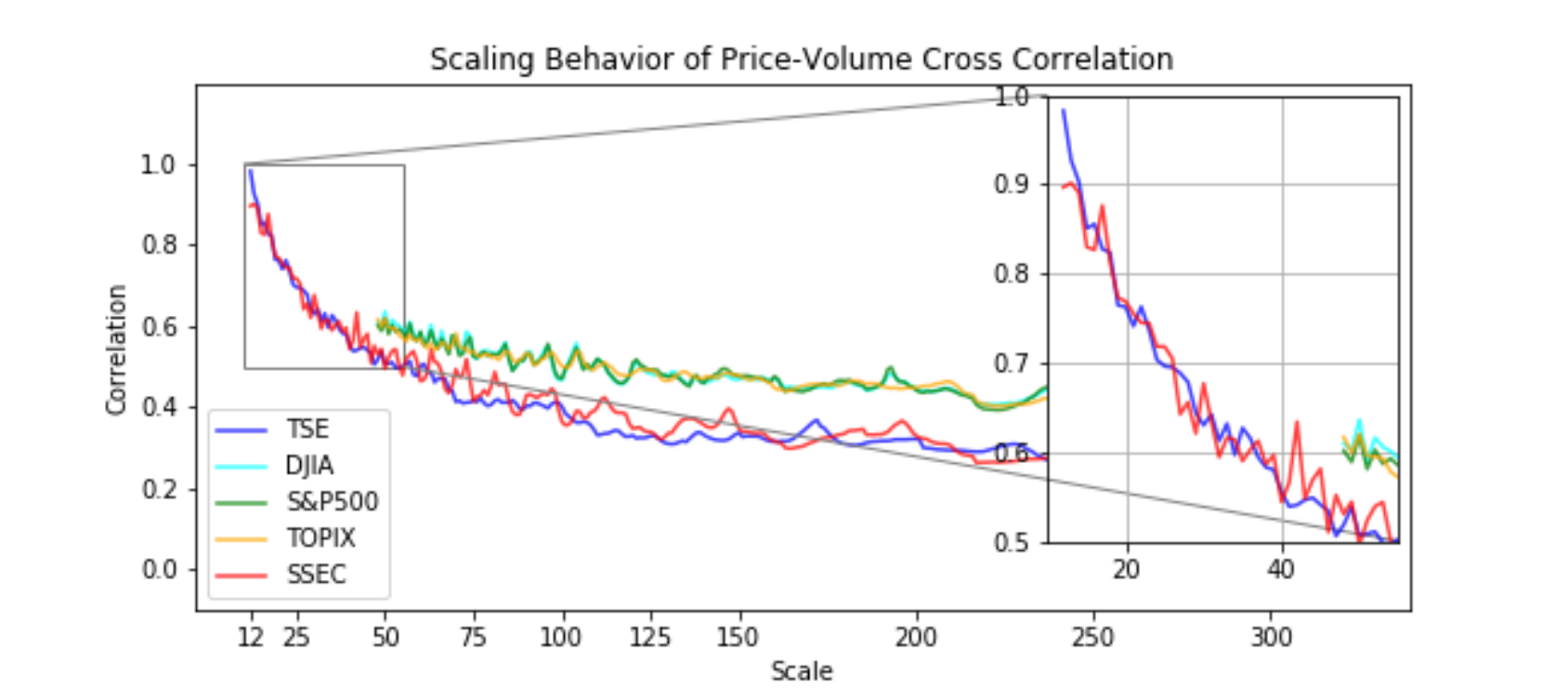}  
	\caption{\textit{Top:} Cross correlation statistics of the investigated
		market versus critical value are shown. \textit{Bottom:} Scaling behavior of
		correlation coefficients of the investigated market are demonstrated.}
	\label{correlation_stats}
\end{figure}

In Fig.\ref{fig_MF}, price multifractalities of all investigated markets are
larger than volume multifractalities of the corresponding markets. If the
time series is multifractal and $q$ is extremely large for the system, it
may yield to $f_{(\alpha )}<0$~\cite{Mandelbrot1990}. On the other hand, $%
f_{(\alpha )}<0$, may occur on extremely large or small singularity spectra~%
\cite{Jiang2007}. Since a negative dimension is an unreal solution, this may warn us that the timeseries histogram is spars-- which is a characteristic of real-world financial timeseries due to the fact that they may not be Guassian. Hence, those events near the tails of the histogram increase and eliminate limited-length effects. In addition, the right panels of Fig.~\ref{fig_MF}, which relate to the
nonlinearity of scaling exponents, prove that price multifractalities occur
more often than volume mulifractalities of the markets we investigated. As a
result, the multifractality of price-volume coupling can be found to be
between the corresponding price and trading volume multifractalities.\newline
As shown in Fig.~\ref{fig_MFcross}, among the investigated markets, TSE has
the least multifractality degree of price-volume coupling. Conversely, for
the developed markets, the multifractality degrees of price-volume coupling
are large than that of emerging markets. However,
considering the singularity spectrum, low efficiency of the markets
contributes to high singularity strength in price-volume couplings.
Furthermore, low efficiency in the markets coincides with a scaling exponent
of price-volume coupling with low curvature. It is notable
that, the price-volume coupling of the developed markets such as DJIA,
S\&P500 and TOPIX (with $H^{price}<0.5$), are highly affected by trading
volume. Hence, their price-volume couplings are not a random
walk (on the contrary to the TSE). 
What is noteworthy, is that the singularity spectrum for an emerging market
such as TSE (with $H^{price}>0.5$) is more left-hooked than that of the
developed ones. The price-volume cross correlation of an emerging market is
more led by some attributes other than volume of transactions (such as the
manipulation of the supply and demand at certain prices which causes markets
to be led more by price rather than trading volume!). This is proven by the
left panels of Fig.\ref{fig_MF}.
\paragraph{Cross Correlation Coefficient}
Firstly, by applying Eq.~\ref{eq13} the scales with significant cross
correlation coefficients are distinguished, as shown in Fig.~\ref%
{correlation_stats}. After a statistical confirmation of correlation
significance, by applying Eq.~\ref{eq14}, we evaluate the scaling behavior
of correlation coefficients of the investigated markets throughout different
time-scales, Fig.~\ref{correlation_stats}, bottom.

We filtered the area with significant cross correlations in order to
investigate the cross correlation coefficients in the next step. Fig.~\ref%
{correlation_stats} contains cross correlation statistics versus critical
value, and Fig.~\ref{correlation_stats}, bottom, shows the corresponding
correlation coefficients in a multiscale pattern--all for $h_{(q=2)}$-- of
price-volume coupling for investigated markets. As shown, the cross
correlation statistics are significant for the studied range of time-scales.
As illustrated in Fig.~\ref{correlation_stats}, bottom, the large temporal
segments size contribute to the small cross-correlation coefficients and
large fluctuation functions (such as ${DFA}_{price}$, ${DFA}_{volume}$, ${DXA%
}_{price-volume}$). When it comes to the large scales, the multifractal
behavior of fluctuation functions corresponding to price and volume, does
not present an extreme correlation and the scaling behavior of price and
trading volumes are totally different in relatively large scales. Hence, in
large scales, it makes no sense to describe price volatilities by trading
volume volatilities. It is noteworthy that, in Fig.~\ref{correlation_stats},
bottom, the price-volume cross correlations of domains maintaining a power
law in both price time series and volume time series in each market, will
show that the decrease of the correlation coefficient co-occurs faster by an
increase of time-scales in the emerging markets such as TSE and SSEC %
rather than that of the developed ones. Consequently, as
shown in Fig.~\ref{correlation_stats} and Fig.~\ref{correlation_stats},
bottom, by increasing time-scale, the behavior of emerging markets deviate
from the behavior of developed markets. 
In the period of our study, it is obvious that among all investigated
markets, the scaling correlation coefficients of the developed markets
maintain to 0.4 to around 0.6. On the other hand, the scaling correlation
coefficients of the emerging markets maintain around less than 1 and
ultimately tend to somehow 0.2 to 0.35 in large time-scales. It means that the decrease in scaling correlation coefficients of emerging
markets is more sensitive rather than that of the developed markets toward
time-scale size.
\section{Conclusion}
In multifractal time series, we need to consider several statistical moments
to describe the behavior of the system. Therefore, investors need to execute
a multifractal analysis to be more familiar with the system's behavior.
Nevertheless, in the aforementioned markets, the generalized Hurst exponent, 
$h_{(q)}$, is a function of q. By illustrating the
singularity spectrum, this study shows nonlinear behavior of price, volume,
and price-volume structure. As a consequence, studying a single variable
without considering simultaneous collective effects and their cross effects,
may be biased.

Since the Hurst exponent of price is larger than 0.5 ($h_{(q=2)}^{price}>0.5$%
) for TSE, it is an emerging market with persistent behavior of price
volatilities. Conversely, because of $h_{(q=2)}^{price}<0.5$, DJIA, S\&P500
and TOPIX, are classified as developed markets with short timescale and anti-persistent behavior. The investigated markets are affected by their memory. Hence, the models which analyze these markets based on the Efficient Market Hypothesis may no longer estimate the market behavior accurately. In
an inefficient market, the short timescale and large timescale phenomena exist in price.\\
Long-range timescales leads to more predictability for price. Long-range behaviors cause more persistent information effects and lessen the speed of fading-out information from past dynamics.
	Furthermore, the trading volume volatilities
	present negatively correlated behavior. Furthermore, the multifractal
	behavior of volume DFA is less than that of price DFA.\newline
Moreover, the cross correlation coefficients show scaling behaviors which
are totally significant in the investigated time-scales and decreases by
increasing time-scales.

Since for the price-volume coupling we have $0.5>H^{TSE}>H^{SSEC}>H^{developed}$, 
it can be concluded that the price-volume couplings for the investigated developed markets 
are significantly negatively correlated and they possess significant valid information.
As a whole, in emerging markets, market behavior is guided 
more by a phenomenon other than trading volume rather than just trading volume.%

Multifractal volatilities in financial markets have become
highly important in risk management. Efficient risk management requires
understanding of information translation between couplings of stock markets
structures and internal and external dynamics. One of the most practical
measures of this study is applying the cross correlation matrix based on
Laloux \textit{et al.}~\cite{Laloux1999} and Plerou \textit{et al.}~\cite%
{Plerou2002} for several fluctuation functions to measure market risk.

\bibliographystyle{elsarticle-num}
\bibliography{mfdxa.bib}

\begin{thebibliography}{10}
\expandafter\ifx\csname url\endcsname\relax
  \def\url#1{\texttt{#1}}\fi
\expandafter\ifx\csname urlprefix\endcsname\relax\def\urlprefix{URL }\fi
\expandafter\ifx\csname href\endcsname\relax
  \def\href#1#2{#2} \def\path#1{#1}\fi

\bibitem{Lux2002}
T.~Lux, M.~Ausloos, Market fluctuations i: Scaling, multiscaling, and their
  possible origins, in: The Science of Disasters, Springer Berlin Heidelberg,
  2002, pp. 372--409.
\newblock \href {http://dx.doi.org/10.1007/978-3-642-56257-0_13}
  {\path{doi:10.1007/978-3-642-56257-0_13}}.

\bibitem{Peng1994}
C.-K. Peng, S.~V. Buldyrev, S.~Havlin, M.~Simons, H.~E. Stanley, A.~L.
  Goldberger, Mosaic organization of {DNA} nucleotides, Physical Review E
  49~(2) (1994) 1685--1689.
\newblock \href {http://dx.doi.org/10.1103/physreve.49.1685}
  {\path{doi:10.1103/physreve.49.1685}}.

\bibitem{Peng1995}
C.-K. Peng, S.~Havlin, H.~E. Stanley, A.~L. Goldberger, Quantification of
  scaling exponents and crossover phenomena in nonstationary heartbeat time
  series, Chaos: An Interdisciplinary Journal of Nonlinear Science 5~(1) (1995)
  82--87.
\newblock \href {http://dx.doi.org/10.1063/1.166141}
  {\path{doi:10.1063/1.166141}}.

\bibitem{Shadkhoo2009}
S.~Shadkhoo, G.~R. Jafari, Multifractal detrended cross-correlation analysis of
  temporal and spatial seismic data, The European Physical Journal B 72~(4)
  (2009) 679--683.
\newblock \href {http://dx.doi.org/10.1140/epjb/e2009-00402-2}
  {\path{doi:10.1140/epjb/e2009-00402-2}}.

\bibitem{Hedayatifar2011}
L.~Hedayatifar, M.~Vahabi, G.~R. Jafari, Coupling detrended fluctuation
  analysis for analyzing coupled nonstationary signals, Physical Review E
  84~(2) (2011) 021138.
\newblock \href {http://dx.doi.org/10.1103/physreve.84.021138}
  {\path{doi:10.1103/physreve.84.021138}}.

\bibitem{Ossadnik1994}
S.~Ossadnik, S.~Buldyrev, A.~Goldberger, S.~Havlin, R.~Mantegna, C.~Peng,
  M.~Simons, H.~Stanley, Correlation approach to identify coding regions in
  {DNA} sequences, Biophysical Journal 67~(1) (1994) 64--70.
\newblock \href {http://dx.doi.org/10.1016/s0006-3495(94)80455-2}
  {\path{doi:10.1016/s0006-3495(94)80455-2}}.

\bibitem{Kantelhardt2002}
J.~W. Kantelhardt, S.~A. Zschiegner, E.~Koscielny-Bunde, S.~Havlin, A.~Bunde,
  H.~Stanley, Multifractal detrended fluctuation analysis of nonstationary time
  series, Physica A: Statistical Mechanics and its Applications 316~(1-4)
  (2002) 87--114.
\newblock \href {http://dx.doi.org/10.1016/s0378-4371(02)01383-3}
  {\path{doi:10.1016/s0378-4371(02)01383-3}}.

\bibitem{Kavasseri2005}
R.~Kavasseri, R.~Nagarajan, A multifractal description of wind speed records,
  Chaos, Solitons {\&} Fractals 24~(1) (2005) 165--173.
\newblock \href {http://dx.doi.org/10.1016/s0960-0779(04)00533-8}
  {\path{doi:10.1016/s0960-0779(04)00533-8}}.

\bibitem{Jiang2007}
Z.-Q. Jiang, W.-X. Zhou, Scale invariant distribution and multifractality of
  volatility multipliers in stock markets, Physica A: Statistical Mechanics and
  its Applications 381 (2007) 343--350.
\newblock \href {http://dx.doi.org/10.1016/j.physa.2007.03.015}
  {\path{doi:10.1016/j.physa.2007.03.015}}.

\bibitem{Podobnik2008}
B.~Podobnik, H.~E. Stanley, Detrended cross-correlation analysis: A new method
  for analyzing two nonstationary time series, Physical Review Letters 100~(8)
  (2008) 084102.
\newblock \href {http://dx.doi.org/10.1103/physrevlett.100.084102}
  {\path{doi:10.1103/physrevlett.100.084102}}.

\bibitem{Podobnik2009}
B.~Podobnik, D.~Horvatic, A.~M. Petersen, H.~E. Stanley, Cross-correlations
  between volume change and price change, Proceedings of the National Academy
  of Sciences 106~(52) (2009) 22079--22084.
\newblock \href {http://dx.doi.org/10.1073/pnas.0911983106}
  {\path{doi:10.1073/pnas.0911983106}}.

\bibitem{Lin2014}
A.~Lin, P.~Shang, H.~Zhou, Cross-correlations and structures of stock markets
  based on multiscale {MF}-{DXA} and {PCA}, Nonlinear Dynamics 78~(1) (2014)
  485--494.
\newblock \href {http://dx.doi.org/10.1007/s11071-014-1455-5}
  {\path{doi:10.1007/s11071-014-1455-5}}.

\bibitem{Sornette2018}
Z.-Q. Jiang, W.-J. Xie, W.-X. Zhou, D.~Sornette, Multifractal analysis of
  financial markets, arXiv preprint arXiv:1805.04750.

\bibitem{Caraiani2015}
P.~Caraiani, E.~Haven, Evidence of multifractality from {CEE} exchange rates
  against euro, Physica A: Statistical Mechanics and its Applications 419
  (2015) 395--407.
\newblock \href {http://dx.doi.org/10.1016/j.physa.2014.06.043}
  {\path{doi:10.1016/j.physa.2014.06.043}}.

\bibitem{Zhuang2014}
X.~Zhuang, Y.~Wei, B.~Zhang, Multifractal detrended cross-correlation analysis
  of carbon and crude oil markets, Physica A: Statistical Mechanics and its
  Applications 399 (2014) 113--125.
\newblock \href {http://dx.doi.org/10.1016/j.physa.2013.12.048}
  {\path{doi:10.1016/j.physa.2013.12.048}}.

\bibitem{Drod2018}
S.~Dro{\.{z}}d{\.{z}}, R.~Gȩbarowski, L.~Minati, P.~O{\'{s}}wiȩcimka,
  M.~Wactorek, Bitcoin market route to maturity? evidence from return
  fluctuations, temporal correlations and multiscaling effects, Chaos: An
  Interdisciplinary Journal of Nonlinear Science 28~(7) (2018) 071101.
\newblock \href {http://dx.doi.org/10.1063/1.5036517}
  {\path{doi:10.1063/1.5036517}}.

\bibitem{Safdari2016}
H.~Safdari, A.~Hosseiny, S.~V. Farahani, G.~Jafari, A picture for the coupling
  of unemployment and inflation, Physica A: Statistical Mechanics and its
  Applications 444 (2016) 744--750.
\newblock \href {http://dx.doi.org/10.1016/j.physa.2015.10.072}
  {\path{doi:10.1016/j.physa.2015.10.072}}.

\bibitem{Torre2017}
S.~Rendón de~la Torre, J.~Kalda, R.~Kitt, J.~Engelbrecht, Fractal and
  multifractal analysis of complex networks: Estonian network of payments, The
  European Physical Journal B 90~(12) (2017) 90--234.
\newblock \href {http://dx.doi.org/10.1140/epjb/e2017-80214-5}
  {\path{doi:10.1140/epjb/e2017-80214-5}}.

\bibitem{Ausloos2004}
M.~Ausloos, Statistical physics in meteorology, Physica A: Statistical
  Mechanics and its Applications 336~(1-2) (2004) 93--101.
\newblock \href {http://dx.doi.org/10.1016/j.physa.2004.01.014}
  {\path{doi:10.1016/j.physa.2004.01.014}}.

\bibitem{ivanova2001multifractality}
K.~Ivanova, N.~Gospodinova, H.~N. Shirer, T.~P. Ackerman, M.~A. Mikhalev,
  M.~Ausloos, Multifractality of cloud base height profiles (2001).
\newblock \href {http://arxiv.org/abs/cond-mat/0108395}
  {\path{arXiv:cond-mat/0108395}}.

\bibitem{Ausloos2012}
M.~Ausloos, Generalized hurst exponent and multifractal function of original
  and translated texts mapped into frequency and length time series, Phys. Rev.
  E 86 (2012) 031108.
\newblock \href {http://dx.doi.org/10.1103/PhysRevE.86.031108}
  {\path{doi:10.1103/PhysRevE.86.031108}}.

\bibitem{Ausloos2017}
M.~Ausloos, R.~Cerqueti, C.~Lupi, Long-range properties and data validity for
  hydrogeological time series: The case of the paglia river, Physica A:
  Statistical Mechanics and its Applications 470 (2017) 39--50.
\newblock \href {http://dx.doi.org/10.1016/j.physa.2016.11.137}
  {\path{doi:10.1016/j.physa.2016.11.137}}.

\bibitem{Movahed2006}
M.~S. Movahed, G.~R. Jafari, F.~Ghasemi, S.~Rahvar, M.~R.~R. Tabar,
  Multifractal detrended fluctuation analysis of sunspot time series, Journal
  of Statistical Mechanics: Theory and Experiment 2006~(02) (2006)
  P02003--P02003.
\newblock \href {http://dx.doi.org/10.1088/1742-5468/2006/02/p02003}
  {\path{doi:10.1088/1742-5468/2006/02/p02003}}.

\bibitem{Ausloos2002}
M.~Ausloos, K.~Ivanova, Mechanistic approach to generalized technical analysis
  of share prices and stock market indices, The European Physical Journal B -
  Condensed Matter 27~(2) (2002) 177--187.
\newblock \href {http://dx.doi.org/10.1140/epjb/e20020144}
  {\path{doi:10.1140/epjb/e20020144}}.

\bibitem{Nasiri2018}
S.~Nasiri, E.~Bektas, G.~Jafari, The impact of trading volume on the stock
  market credibility: Bohmian quantum potential approach, Physica A:
  Statistical Mechanics and its Applications 512 (2018) 1104--1112.
\newblock \href {http://dx.doi.org/10.1016/j.physa.2018.08.026}
  {\path{doi:10.1016/j.physa.2018.08.026}}.

\bibitem{Guo2012}
Y.~Guo, J.~Huang, H.~Cheng, Multifractal features of metal futures market based
  on multifractal detrended cross-correlation analysis 41~(10) (2012)
  1509--1525.
\newblock \href {http://dx.doi.org/10.1108/03684921211276710}
  {\path{doi:10.1108/03684921211276710}}.

\bibitem{Chen2001}
G.~meng Chen, M.~Firth, O.~M. Rui, The dynamic relation between stock returns,
  trading volume, and volatility, The Financial Review 36~(3) (2001) 153--174.
\newblock \href {http://dx.doi.org/10.1111/j.1540-6288.2001.tb00024.x}
  {\path{doi:10.1111/j.1540-6288.2001.tb00024.x}}.

\bibitem{Chuang2009}
C.-C. Chuang, C.-M. Kuan, H.-Y. Lin, Causality in quantiles and dynamic stock
  return{\textendash}volume relations, Journal of Banking {\&} Finance 33~(7)
  (2009) 1351--1360.
\newblock \href {http://dx.doi.org/10.1016/j.jbankfin.2009.02.013}
  {\path{doi:10.1016/j.jbankfin.2009.02.013}}.

\bibitem{Campbell1993}
J.~Y. Campbell, S.~J. Grossman, J.~Wang, Trading volume and serial correlation
  in stock returns, The Quarterly Journal of Economics 108~(4) (1993) 905--939.
\newblock \href {http://dx.doi.org/10.2307/2118454}
  {\path{doi:10.2307/2118454}}.

\bibitem{Ahmad2016}
M.~Ahmad, A.~Sarr, Joint distribution of stock market returns and trading
  volume, Rev. Integr. Bus. Econ. Res. 5 (2016) 110--116.
\newblock \href
  {http://dx.doi.org/http://sibresearch.org/uploads/3/4/0/9/34097180/riber_b16-085_110-116.pdf}
  {\path{doi:http://sibresearch.org/uploads/3/4/0/9/34097180/riber_b16-085_110-116.pdf}}.

\bibitem{Osborne1959}
M.~F.~M. Osborne, \href{http://www.jstor.org/stable/167153}{Brownian motion in
  the stock market}, Operations Research 7~(2) (1959) 145--173.
\newline\urlprefix\url{http://www.jstor.org/stable/167153}

\bibitem{Saatcioglu1998}
K.~Saatcioglu, L.~T. Starks, The stock price{\textendash}volume relationship in
  emerging stock markets: the case of latin america, International Journal of
  Forecasting 14~(2) (1998) 215--225.
\newblock \href {http://dx.doi.org/10.1016/s0169-2070(98)00028-4}
  {\path{doi:10.1016/s0169-2070(98)00028-4}}.

\bibitem{Wang1994}
J.~Wang, A model of competitive stock trading volume, Journal of Political
  Economy 102~(1) (1994) 127--168.
\newblock \href {http://dx.doi.org/10.1086/261924} {\path{doi:10.1086/261924}}.

\bibitem{Huang2018}
W.~Huang, K.~Mazouz, Excess cash, trading continuity, and liquidity risk,
  Journal of Corporate Finance 48 (2018) 275--291.
\newblock \href {http://dx.doi.org/10.1016/j.jcorpfin.2017.11.005}
  {\path{doi:10.1016/j.jcorpfin.2017.11.005}}.

\bibitem{Buldyrev1995}
S.~Buldyrev, A.~Goldberger, S.~Havlin, R.~N. Mantegna, M.~Matsa, C.~Peng,
  M.~Simons, H.~Stanley, Long-range correlation properties of coding and
  noncoding dna sequences: Genbank analysis, Phys. Rev. E 51 (1995) 5084--5091.

\bibitem{Barnes1966}
J.~Barnes, D.~Allan, A statistical model of flicker noise, Proceedings of the
  IEEE 54~(2) (1966) 176--178.
\newblock \href {http://dx.doi.org/10.1109/proc.1966.4630}
  {\path{doi:10.1109/proc.1966.4630}}.

\bibitem{TAQQU1995}
M.~S. Taqqu, V.~Teverovsky, W.~Willinger, {ESTIMATORS} {FOR} {LONG}-{RANGE}
  {DEPENDENCE}: {AN} {EMPIRICAL} {STUDY}, Fractals 03~(04) (1995) 785--798.
\newblock \href {http://dx.doi.org/10.1142/s0218348x95000692}
  {\path{doi:10.1142/s0218348x95000692}}.

\bibitem{Kantelhardt2001}
J.~W. Kantelhardt, E.~Koscielny-Bunde, H.~H. Rego, S.~Havlin, A.~Bunde,
  Detecting long-range correlations with detrended fluctuation analysis,
  Physica A: Statistical Mechanics and its Applications 295~(3-4) (2001)
  441--454.
\newblock \href {http://dx.doi.org/10.1016/s0378-4371(01)00144-3}
  {\path{doi:10.1016/s0378-4371(01)00144-3}}.

\bibitem{Hu2001}
K.~Hu, P.~C. Ivanov, Z.~Chen, P.~Carpena, H.~E. Stanley, Effect of trends on
  detrended fluctuation analysis, Physical Review E 64~(1) (2001) 011114.
\newblock \href {http://dx.doi.org/10.1103/physreve.64.011114}
  {\path{doi:10.1103/physreve.64.011114}}.

\bibitem{Chen2002}
Z.~Chen, P.~C. Ivanov, K.~Hu, H.~E. Stanley, Effect of nonstationarities on
  detrended fluctuation analysis, Physical Review E 65~(4) (2002) 041107.
\newblock \href {http://dx.doi.org/10.1103/physreve.65.041107}
  {\path{doi:10.1103/physreve.65.041107}}.

\bibitem{Kwapie2012}
J.~Kwapie{\'{n}}, S.~Dro{\.{z}}d{\.{z}}, Physical approach to complex systems,
  Physics Reports 515~(3-4) (2012) 115--226.
\newblock \href {http://dx.doi.org/10.1016/j.physrep.2012.01.007}
  {\path{doi:10.1016/j.physrep.2012.01.007}}.

\bibitem{Chen2016}
H.-Y. Chen, C.-F. Lee, W.~K. Shih, Technical, fundamental, and combined
  information for separating winners from losers, Pacific-Basin Finance Journal
  39 (2016) 224--242.
\newblock \href {http://dx.doi.org/10.1016/j.pacfin.2016.06.008}
  {\path{doi:10.1016/j.pacfin.2016.06.008}}.

\bibitem{Pece2015}
A.~M. Pece, N.~Petria, Volatility, thin trading and non-liniarities: An
  empirical approach for the {BET} index during pre-crisis and post-crisis
  periods, Procedia Economics and Finance 32 (2015) 1342--1352.
\newblock \href {http://dx.doi.org/10.1016/s2212-5671(15)01511-7}
  {\path{doi:10.1016/s2212-5671(15)01511-7}}.

\bibitem{Ozturk2017}
S.~R. Ozturk, M.~van~der Wel, D.~van Dijk, Intraday price discovery in
  fragmented markets, Journal of Financial Markets 32 (2017) 28--48.
\newblock \href {http://dx.doi.org/10.1016/j.finmar.2016.10.001}
  {\path{doi:10.1016/j.finmar.2016.10.001}}.

\bibitem{Hong2016}
K.~Hong, E.~Wu, The roles of past returns and firm fundamentals in driving us
  stock price movements, International Review of Financial Analysis 43~(C)
  (2016) 62--75.

\bibitem{Chan2015}
H.~Chan, P.~Docherty, Momentum in australian style portfolios: risk or
  inefficiency?, Accounting {\&} Finance 56~(2) (2015) 333--361.
\newblock \href {http://dx.doi.org/10.1111/acfi.12106}
  {\path{doi:10.1111/acfi.12106}}.

\bibitem{Dash2016}
R.~Dash, P.~K. Dash, A hybrid stock trading framework integrating technical
  analysis with machine learning techniques, The Journal of Finance and Data
  Science 2~(1) (2016) 42--57.
\newblock \href {http://dx.doi.org/10.1016/j.jfds.2016.03.002}
  {\path{doi:10.1016/j.jfds.2016.03.002}}.

\bibitem{Abarbanell1997}
J.~S. Abarbanell, B.~J. Bushee, Fundamental analysis, future earnings, and
  stock prices, Journal of Accounting Research 35~(1) (1997) 1.
\newblock \href {http://dx.doi.org/10.2307/2491464}
  {\path{doi:10.2307/2491464}}.

\bibitem{Jamali2015}
T.~Jamali, G.~R. Jafari, Spectra of empirical autocorrelation matrices: A
  random-matrix-theory{\textendash}inspired perspective, {EPL} (Europhysics
  Letters) 111~(1) (2015) 10001.
\newblock \href {http://dx.doi.org/10.1209/0295-5075/111/10001}
  {\path{doi:10.1209/0295-5075/111/10001}}.

\bibitem{Tahmasebi2015}
F.~Tahmasebi, S.~Meskinimood, A.~Namaki, S.~V. Farahani, S.~Jalalzadeh, G.~R.
  Jafari, Financial market images: A practical approach owing to the secret
  quantum potential, {EPL} (Europhysics Letters) 109~(3) (2015) 30001.
\newblock \href {http://dx.doi.org/10.1209/0295-5075/109/30001}
  {\path{doi:10.1209/0295-5075/109/30001}}.

\bibitem{Podobnik2011}
B.~Podobnik, Z.-Q. Jiang, W.-X. Zhou, H.~E. Stanley, Statistical tests for
  power-law cross-correlated processes, Physical Review E 84~(6) (2011) 066118.
\newblock \href {http://dx.doi.org/10.1103/physreve.84.066118}
  {\path{doi:10.1103/physreve.84.066118}}.

\bibitem{Zhu2018}
H.~Zhu, W.~Zhang, Multifractal property of chinese stock market in the csi 800
  index based on mf-dfa approach, Physica A: Statistical Mechanics and its
  Applications 490 (2018) 497--503.
\newblock \href {http://dx.doi.org/10.1016/j.physa.2017.08.060}
  {\path{doi:10.1016/j.physa.2017.08.060}}.

\bibitem{Carbone2004}
A.~Carbone, G.~Castelli, H.~Stanley, Time-dependent hurst exponent in financial
  time series, Physica A: Statistical Mechanics and its Applications 344~(1-2)
  (2004) 267--271.
\newblock \href {http://dx.doi.org/10.1016/j.physa.2004.06.130}
  {\path{doi:10.1016/j.physa.2004.06.130}}.

\bibitem{Mantegna2000}
R.~N. Mantegna, H.~E. Stanley, An Introduction to Econophysics: Correlations
  and Complexity in Finance, Cambridge: Cambridge University Press, 2000.

\bibitem{Liu1999}
Y.~Liu, P.~Gopikrishnan, Cizeau, Meyer, Peng, H.~E. Stanley, Statistical
  properties of the volatility of price fluctuations, Physical Review E 60~(2)
  (1999) 1390--1400.
\newblock \href {http://dx.doi.org/10.1103/physreve.60.1390}
  {\path{doi:10.1103/physreve.60.1390}}.

\bibitem{Vandewalle1999}
N.~Vandewalle, M.~Ausloos, P.~Boveroux, A.~Minguet, Visualizing the
  log-periodic pattern before crashes, The European Physical Journal B 9~(2)
  (1999) 355--359.
\newblock \href {http://dx.doi.org/10.1007/s100510050775}
  {\path{doi:10.1007/s100510050775}}.

\bibitem{Yin2013}
Y.~Yin, P.~Shang, Modified {DFA} and {DCCA} approach for quantifying the
  multiscale correlation structure of financial markets, Physica A: Statistical
  Mechanics and its Applications 392~(24) (2013) 6442--6457.
\newblock \href {http://dx.doi.org/10.1016/j.physa.2013.07.070}
  {\path{doi:10.1016/j.physa.2013.07.070}}.

\bibitem{Zunino2008}
L.~Zunino, B.~Tabak, A.~Figliola, D.~P{\'{e}}rez, M.~Garavaglia, O.~Rosso, A
  multifractal approach for stock market inefficiency, Physica A: Statistical
  Mechanics and its Applications 387~(26) (2008) 6558--6566.
\newblock \href {http://dx.doi.org/10.1016/j.physa.2008.08.028}
  {\path{doi:10.1016/j.physa.2008.08.028}}.

\bibitem{Zhou2008}
W.-X. Zhou, Multifractal detrended cross-correlation analysis for two
  nonstationary signals, Physical Review E 77~(6) (2008) 066211.
\newblock \href {http://dx.doi.org/10.1103/physreve.77.066211}
  {\path{doi:10.1103/physreve.77.066211}}.

\bibitem{Cottet2004}
A.~Cottet, W.~Belzig, C.~Bruder, Positive cross correlations in a
  three-terminal quantum dot with ferromagnetic contacts, Physical Review
  Letters 92~(20) (2004) 206801.
\newblock \href {http://dx.doi.org/10.1103/physrevlett.92.206801}
  {\path{doi:10.1103/physrevlett.92.206801}}.

\bibitem{Campillo2003}
M.~Campillo, Long-range correlations in the diffuse seismic coda, Science
  299~(5606) (2003) 547--549.
\newblock \href {http://dx.doi.org/10.1126/science.1078551}
  {\path{doi:10.1126/science.1078551}}.

\bibitem{Hajian2010}
S.~Hajian, M.~S. Movahed, Multifractal detrended cross-correlation analysis of
  sunspot numbers and river flow fluctuations, Physica A: Statistical Mechanics
  and its Applications 389~(21) (2010) 4942--4957.
\newblock \href {http://dx.doi.org/10.1016/j.physa.2010.06.025}
  {\path{doi:10.1016/j.physa.2010.06.025}}.

\bibitem{Shayeganfar2009}
F.~Shayeganfar, S.~Jabbari-Farouji, M.~S. Movahed, G.~R. Jafari, M.~R.~R.
  Tabar, Multifractal analysis of light scattering-intensity fluctuations,
  Physical Review E 80~(6).
\newblock \href {http://dx.doi.org/10.1103/physreve.80.061126}
  {\path{doi:10.1103/physreve.80.061126}}.

\bibitem{Schumann2011}
A.~Y. Schumann, J.~W. Kantelhardt, Multifractal moving average analysis and
  test of multifractal model with tuned correlations, Physica A: Statistical
  Mechanics and its Applications 390~(14) (2011) 2637--2654.
\newblock \href {http://dx.doi.org/10.1016/j.physa.2011.03.002}
  {\path{doi:10.1016/j.physa.2011.03.002}}.

\bibitem{Jafari2007}
G.~R. Jafari, P.~Pedram, L.~Hedayatifar, Long-range correlation and
  multifractality in bach's inventions pitches, Journal of Statistical
  Mechanics: Theory and Experiment 2007~(04) (2007) P04012--P04012.
\newblock \href {http://dx.doi.org/10.1088/1742-5468/2007/04/p04012}
  {\path{doi:10.1088/1742-5468/2007/04/p04012}}.

\bibitem{Mandelbrot1990}
B.~B. Mandelbrot, Negative fractal dimensions and multifractals, Physica A:
  Statistical Mechanics and its Applications 163~(1) (1990) 306--315.
\newblock \href {http://dx.doi.org/10.1016/0378-4371(90)90339-t}
  {\path{doi:10.1016/0378-4371(90)90339-t}}.

\bibitem{Laloux1999}
L.~Laloux, P.~Cizeau, J.-P. Bouchaud, M.~Potters, Noise dressing of financial
  correlation matrices, Physical Review Letters 83~(7) (1999) 1467--1470.
\newblock \href {http://dx.doi.org/10.1103/physrevlett.83.1467}
  {\path{doi:10.1103/physrevlett.83.1467}}.

\bibitem{Plerou2002}
V.~Plerou, P.~Gopikrishnan, B.~Rosenow, L.~A.~N. Amaral, T.~Guhr, H.~E.
  Stanley, Random matrix approach to cross correlations in financial data,
  Physical Review E 65~(6) (2002) 066126.
\newblock \href {http://dx.doi.org/10.1103/physreve.65.066126}
  {\path{doi:10.1103/physreve.65.066126}}.

\end{thebibliography}

\end{document}